\documentclass[useAMS,usenatbib]{mn2e} 
\usepackage{times,graphicx,aas_macros,natbib}

\begin{document}

\title[Self-organisation for surveys]{Unsupervised self-organised mapping: a
versatile empirical tool for object selection, classification and redshift
estimation in large surveys} \author[J. E.
Geach]{\parbox[h]{\textwidth}{James\ E.\ Geach\thanks{Banting Fellow; e-mail:\
jimgeach@physics.mcgill.ca}} \vspace*{6pt}\\ \noindent Department of Physics,
McGill University, Ernest Rutherford Building, 3600 rue University,
Montr\'eal, Qu\'ebec, Canada. H3A\ 2T8.}

\pagerange{\pageref{firstpage}--\pageref{lastpage}}\pubyear{2011}

\maketitle

\label{firstpage}

\begin{abstract} We present an application of unsupervised machine learning --
the self-organised map (SOM) -- as a tool for visualising, exploring and
mining the catalogues of large astronomical surveys. Self-organisation
culminates in a low-resolution representation of the `topology' of a parameter
volume, and this can be exploited in various ways pertinent to astronomy.
Using data from the Cosmological Evolution Survey (COSMOS), we demonstrate two
key astronomical applications of the SOM: (i) object classification and
selection, using the example of galaxies with active galactic nuclei as a
demonstration, and (ii) photometric redshift estimation, illustrating how SOMs
can be used as totally empirical predictive tools. With a training set of
$\sim$3800 galaxies with $z_{\rm spec}\leq1$, we achieve photometric redshift
accuracies competitive with other (mainly template fitting) techniques that
use a similar number of photometric bands ($\sigma(\Delta z)=0.03$ with a
$\sim$2\% outlier rate when using {\it u$^*$}-band to 8$\mu$m photometry). We
also test the SOM as a photo-{\it z} tool using the PHoto-{\it z} Accuracy
Testing (PHAT) synthetic catalogue of Hildebrandt et al.\ (2010), which
compares several different photo-{\it z} codes using a common input/training
set. We find that the SOM can deliver accuracies that are competitive with
many of the established template-fitting and empirical methods. This technique
is not without clear limitations, which are discussed, but we suggest it could
be a powerful tool in the era of extremely large -- `petabyte' -- databases
where efficient data-mining is a paramount concern. \end{abstract}

\section{Introduction}

Extremely large surveys provide the means to take great steps forward in a
wide range of astronomical fields, because they probe the large volumes
required to detect those very rare objects that would otherwise be nearly
impossible to find, and yield the immense sample sizes essential for robust
statistical analyses. The crowning achievement of such an approach has
undoubtably been the Sloan Digital Sky Survey (SDSS, York et al.\ 2000), which
at the time of writing is in its eighth data release, and now covers 14\,000
square degrees of imaging, and has obtained spectra for millions of objects
(Aihara et al.\ 2011). SDSS marked the beginning of an era of extremely large
digital sky surveys and continues to demonstrate its power (in the form of
`SDSS-III') across a remarkably wide range of scientific areas, from Galactic
studies to cosmology. The Panoramic Survey Telescope and Rapid Response System
(Pan-STARRS), Large Synoptic Survey Telescope (LSST) and Dark Energy Survey
(DES), amongst others are poised to take up the mantle set by SDSS in the last
decade. These surveys will provide deeper optical and near-infrared imaging
over the majority of the solid angle of the sky.

The coming years and decades will see this panoramic approach spill over into
other frequency domains; indeed, our view of the Universe already being
transformed by more sensitive large area surveys in the infrared and
submillimeter (e.g.\ {\it Wide-Field Infrared Explorer} [Wright et al.\ 2010],
{\it Herschel} [e.g.\ Eales et al.\ 2010], the Submillimeter Common User
Bolomoter Array--2) and soon the radio regimes (e.g.\ LOw Frequency ARray
[LOFAR], and Square Kilometer Array pathfinders [e.g.\ Norris et al.\ 2011]).
We are certainly entering exciting times in terms of our capability to survey
the Universe across most of the electromagnetic spectrum, but these large
surveys pose a common challenge: how does one efficiently mine the parameter
volume when we move into the petabyte regime of information content?
Innovative techniques that can efficiently sift and filter the myriad data
will be vital, since often one is interested in selecting for a very specific
subset of data, for example searching for rare objects (populations of
high-redshift galaxies, quasars, low-mass stars or gravitational lenses for
example) or events (supernovae, gamma-ray bursts and other transient
phenomena).

By design, most of the next-generation panoramic surveys will involve simple
continuum imaging. However, many projects will be complemented by ancillary
sub-surveys that will accrue more (deeper) imaging and spectroscopic
observations over smaller areas. The largest surveys that cover most of the
sky will necessarily overlap with fields that have already undergone
significant observational investments. Several of these (e.g.\ the Great
Observatories Origins Deep Survey, Chandra Deep Fields, Cosmological Evolution
Survey [COSMOS] field and the like) have been targeted with deep imaging from
virtually all terrestrial and space-based facilities, and often have been
subject to extensive spectroscopic campaigns, providing very large, deep
redshift catalogues (e.g.\ Lilly et al.\ 2007). As time goes on, the interplay
between overlapping surveys will become more important as we move towards a
truly holistic picture of the sky.

Artificial neural networks, or more generally the technique of `machine
learning', has been used as a tool in astronomy (as well as other scientific
disciplines and industrial applications) for some time (e.g.\ Storrie-Lombardi
et al.\ 1992; Lahav et al.\ 1995); the estimation of galaxy redshifts from
photometry is a classic example. Perhaps the most successful application of
neural networks in an astronomical setting in the past few years has been the
`crowd sourcing' technique of Galaxy Zoo (Raddick et al.\ 2008; Lintott et
al.\ 2008), which exploits thousands of humans to perform visual
classification of galaxies from the SDSS, relying on the superior capabilities
of the human brain (i.e.\ a {\it real} neural network) for pattern
recognition. Many of the established machine learning techniques employ
supervised learning, where the neural network is trained using a series of
input pairs. A common example is the multi-layer perceptron, where each input
pair consists of a vector (a set of photometry for example) and a known output
(a spectroscopic redshift). The network then attempts to find the mapping that
successfully converts the input vectors to the required output -- in this
sense, it is `supervised' learning. After training, new input vectors (e.g.
photometry) can be passed through the network to predict their output
(redshifts, e.g.\ Collister \& Lahav\ 2004).

An alternative approach is to use the input vectors themselves to find the
mapping, since if such a mapping between parameters (or combinations of
parameters) exists, this information should be latent in the input catalogue.
In this paper, I will describe a specific type of unsupervised machine
learning -- the Kohonen self-organising map (SOM, Kohonen\ 1982, 2001) -- as a
tool for data mining in astronomy. SOMs have found application in other
scientific disciplines, notably geophysics and genetics, and other disparate
areas, especially those where some form of pattern recognition is required.
While there has been some use of self-organisation and SOMs in astronomical
applications (e.g.\ N\'u\~nez \& Llacer\ 2003, Mahdi\ 2011), the technique is
not in widespread use. In essence, a SOM is a neural network that takes as
input a large training set (in this case large astronomical catalogues), and
maps it by a process of competitive learning, where neurons compete to become
more like members of the training set. The resulting map is a representation
of the topology of the input parameter space, encoding correlations between
parameters, and allowing one to visualise the high-dimensional properties of a
parameter volume in a low-resolution, lower-dimensional way.

This self-organised mapping has often been described as a form of non-linear
principle component analysis, and is a variant of {\it k}-means clustering
algorithms. It allows one to identify special features (for instance,
clustering) in the input catalogue. As the algorithm itself is effectively
classifying new input data based on previously seen specimens, after training
SOMs be used to `classify' new inputs, even if they only contain a sub-set of
the original information used to train the original map. The method is
unsupervised in the sense that the user is not required to specify the desired
output, as the `mapping' of components of the input vectors is a natural
outcome of the learning. Indeed, perhaps the most fascinating aspect of large
SOMs is the potential for emergent behaviour (Ultsch\ 2007) allowing one to
discover new properties of the input data that would be imperceptible
otherwise.

In summary, the SOM is an extremely versatile tool, and could have several
possible uses, however in this paper we demonstrate two of its main
applications: object selection and parameter estimation. In \S2 we describe
the algorithm, including a toy example and in \S3 we present the two practical
examples using data from the the COSMOS
field\footnote{http://cosmos.astro.caltech.edu/} (Scoville et al.\ 2007). At
the end of the paper we provide a brief list of common SOM terminology for
convenience. The SOM algorithm used here as a Python class is made available
at http://www.physics.mcgill.ca/$\sim$jimgeach/som, or from the author on
request.

\section{Self organisation}

\subsection{Learning philosophy}

The SOM can be considered as a collection of `nodes' arranged in a grid of
arbitrary dimension, although for visualisation purposes two dimensions are
most common. Each node is attached to a vector of `weights' ${\mathbf w}$ with
the same dimension as an input `training' vector, ${\mathbf t}$. In the case
of a galaxy survey for example, a ${\mathbf t}$ could comprise of five
measurements of {\it ugriz} photometry. In fact, the input data need not
actually be vectorised; any input -- provided it can be digitized -- way could
be considered. A further example to consider might be a digital astronomical
image, where we might expect the SOM to help perform morphological
classifications. Througout this work however, we will consider the case of
inputs that are represented as vectors, with each vector component made up
from standard `catalogue data' such as photometry and redshift information.
The map can be considered as a set of `component planes', with a given node in
the $i^{\rm th}$ plane taking the value of $w_i$. In the two dimensional
representation, plotting two or more component planes next to each other
provides a low-resolution visual representation of the higher-dimensional
topology of the input data.

How does the SOM achieve this mapping? To start, each node is initialised with
a random weight; this can be selected from a uniform distribution, or an
arbitrary probability distribution (limited according to a sensible physical
range of values), or even randomly sampled from the input training set. The
learning process is then a set of iterations, and follows a simple philosophy:
each node `competes' to be the best match to a randomly selected vector from
the training set. The winning node -- called the Best Matching Unit (BMU) --
is rewarded by being allowed to become more like the input vector. In
addition, nodes in the vicinity of the BMU, $r_{\rm BMU}$, are also allowed to
be altered in the same direction, but to a lesser extent than the BMU\footnote{For convenience, we provide a table of SOM nomenclature at the end of the paper.}. 

After many samplings, the nodes can learn to become more like the training
set, with the distribution of weights representing the probability
distribution of the training set and the relationship between the components
of individual weights encoding correlations between parameters. Most
importantly, similar nodes get grouped together in the map. This allows one to
examine the parameter space topology, and can be used to search for clusters
within the parameter space of the training set, and thus provides a means of
object classification. In addition, the BMU of any new test galaxy (for
example) contains the SOM's `best guess' of what that galaxy's parameters
should be, based on similar galaxies it has seen before. In the case of
incomplete data for a new test galaxy (for example a missing redshift), the
BMU can provide a prediction for what that missing parameter should be. Thus,
the SOM can be a predictive tool.

The process of learning occurs over a series of $N$ iterations. At each
iteration $t$, nodes compete to be the best match to a randomly selected
training vector, with the BMU being rewarded by changing its weight vector in
the direction of the training vector. Crucially, nodes within some vicinity of
the BMU ($r<r_{\rm BMU}$) are {\it also} allowed to adapt, but to a lesser
extent than the BMU. The effect is that nodes with similar properties end up
grouped close to each other on the map. The adaptation is set by a learning
handicap, called the `neighbourhood function' $R$ that falls off with $r$, and
decays with learning time. The exact form of the neighbourhood function, $R$,
is arbitrary, but a gaussian function is often chosen as a suitable form:
\begin{equation}R = e^{-r / \sigma } \end{equation} where $\sigma$ depends on
time: \begin{equation}\sigma = r_{\rm BMU}e^{-t/\tau}.\end{equation} Here
$\tau$ is a decay constant, usually chosen to be equal to the number of
iterations, $N$.

The region of influence around the BMU $r_{\rm BMU}$ shrinks over time $t$,
such that ever smaller regions of the SOM are allowed to adapt as
$t\rightarrow N$: \begin{equation}r_{\rm BMU} = r^0_{\rm BMU}(1 - t/N).
\end{equation} where $r^0_{\rm BMU}$ is taken to be half of the size of the
map. Finally, {\it all} nodes in the SOM have their learning handicapped over
time, with an additional factor, \begin{equation}L =
e^{-t/\tau}.\end{equation} The effect of these decaying learning rates and
neighbourhood function is sequence of refinement, where the most dramatic and
coarse organisation of nodes occurs early in the learning process, with
subsequent steps fine-tuning the SOM on smaller scales and resolving more
subtle topology in the data.

\begin{figure*}
\centerline{\includegraphics[scale=0.675,angle=-90]{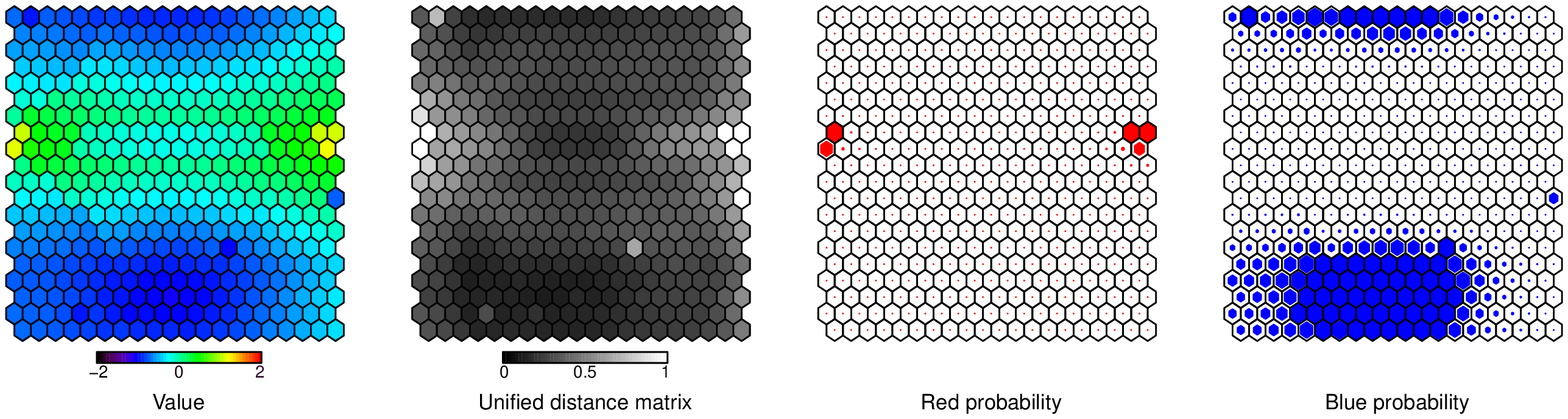}}
\vspace{-0.4cm}
\centerline{\includegraphics[scale=0.675,angle=-90]{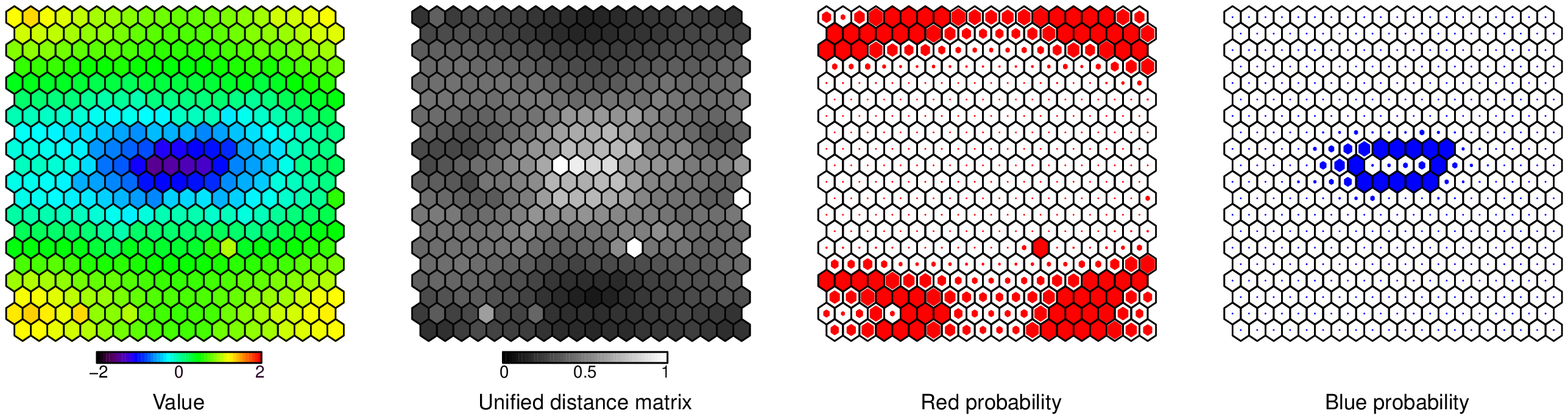}}
\vspace{-0.4cm}
\centerline{\includegraphics[scale=0.675,angle=-90]{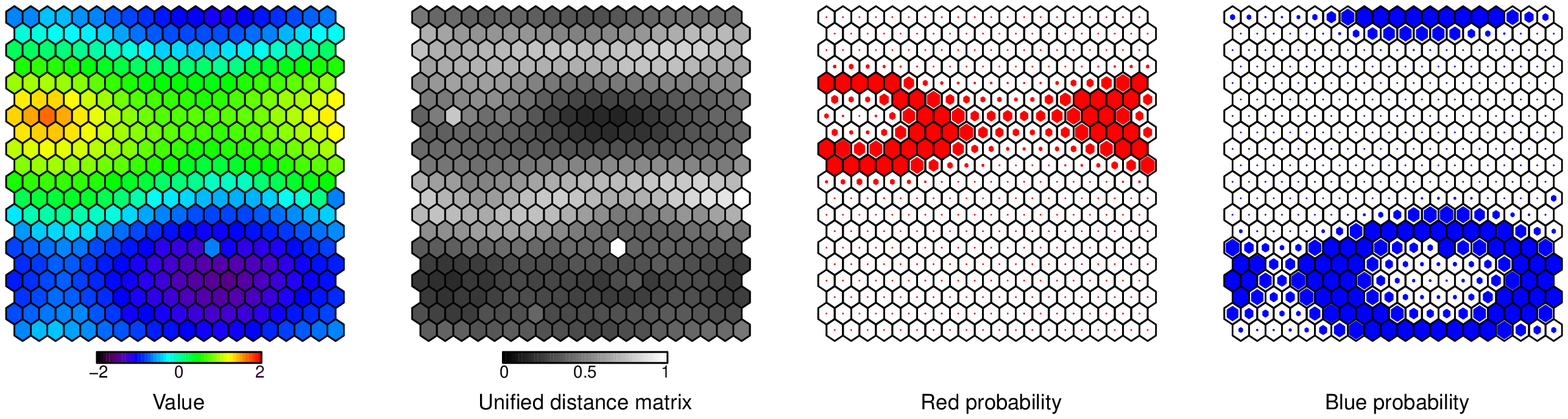}}
\vspace{-0.4cm}
\centerline{\includegraphics[scale=0.675,angle=-90]{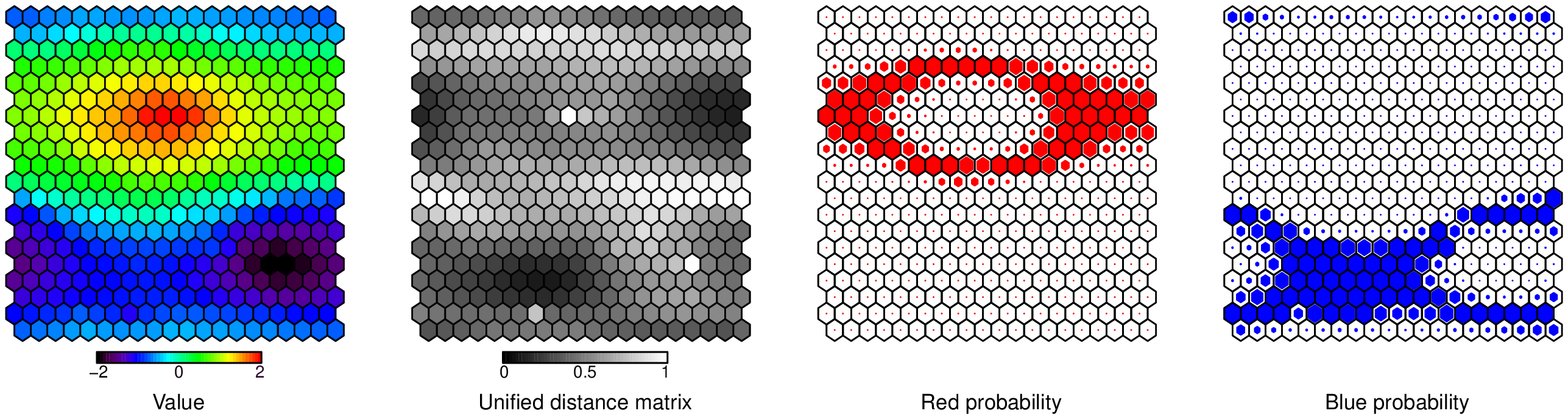}} \caption{Toy
example of self-organisation. We consider two populations, `Red' and `Blue',
defined by two gaussian distributions with means of $1$ and $-1$ respectively.
We generate a catalogue containing 10\,000 of each class, randomly drawing
from the two gaussians, and allow a $20\times20$ node SOM to organise the
values. The different rows show different stages of the learning (1000,
10\,000, 100\,000 and 200\,000 iterations top to bottom). The bottom left
panel shows the clear segregation of map into two distinct regions
representing the two populations; note how the maximum and minimum values are
separated by a large grid distance (the hexagonal cell representation is
traditional for SOM visualisation). Boundaries defining the two clusters are
made apparent by the Unified Distance Matrix, which represents the average
distance (in parameter space) between nodes (a greyscale colour scheme is
traditional for the UDM). Collections of nodes with low UDM values bordered by
swathes with high UDM values can be considered distinct clusters. Finally, the
two right-hand columns show a visualisation where the size of the coloured
nodes are scaled with the probability that the node value is drawn from the
Red or Blue distributions, clearly showing how effective the SOM is at cleanly
selecting the two classes in this low-resolution mapping.} \label{fig:toy}
\end{figure*}

\begin{figure*}
\centerline{\includegraphics[scale=0.7,angle=-90]{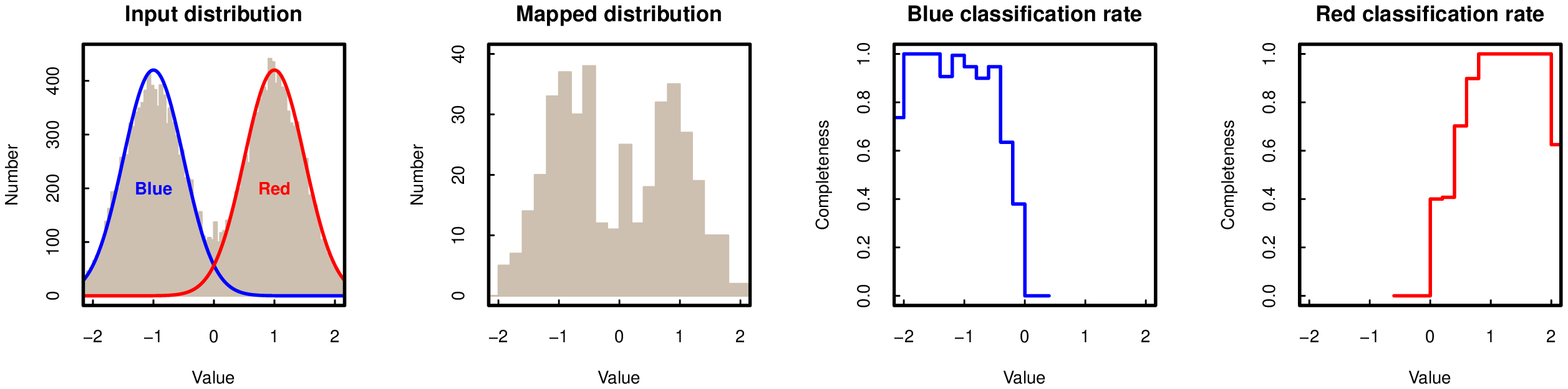}}
\vspace{-0.5cm} \caption{(from left) Input toy training sample of `Red' and
`Blue' objects drawn from two gaussian distributions; distribution of node
weight values in trained SOM (Fig\,1) showing how nodes are
distributed in a way that represents training set. The two right-hand panels
show the fraction of correctly identified Blue and Red objects from a new test
sample of 2000 objects, where we have used the UDM division of the SOM into
two main regions to classify new test data. The SOM can successfully identify
new inputs, and the properties of the recovered distributions are well matched
to the true underlying sample distributions.}\label{fig:toy_dist}
\end{figure*}

\subsection{The algorithm}

From the learning sequence described in \S2.1, the algorithm itself can be
summarised as follows:

\begin{enumerate}

\item[1)] Initialise the SOM by randomly assigning vectors to each node. The
vectors can be selected uniformly from within some suitably limited parameter
volume, or take the values of vectors sampled randomly from the training set.

\item[2)] A training vector $\mathbf t$ is picked randomly from the
training set.

\item[3)] For the $i^{\rm th}$ node in the map described by weight vector
${\bf w}_i$, the Euclidean distance of that weight from the $j^{\rm th}$
training vector ${\bf t}_j$ is assessed: $d_i = |{\bf t}_j - {\bf w}_i|$. The
winning node has $\min(d)$ -- i.e. it was the closest to the input vector, and
becomes the BMU.

\item[4)] Every node within the region of influence, $r_{\rm BMU}$, is
allowed to be pulled in the direction of the input vector, weighted by
the learning factors (equations 1--4): ${\bf w}^\prime_i = {\bf w}_i +
{{\delta}_j}\times R\times L\times({\bf t}_j - {\bf w}_i)$. The factor
${{\delta}_j}$ is an optional additional weighting that can take into
account the measurement uncertainty, or data quality of each element of
the training vector. This penalises unreliable data by not allowing it
to contribute heavily to the development of the map.

\item[5)] Repeat steps 2--4 for $N$ iterations (where $N$ is sufficiently
large that is over-samples the training set several times), or until $r_{\rm
BMU}=1$ (or a user chosen minimum). \end{enumerate}

After many iterations, the SOM will evolve such that similar regions are
geometrically close to each other on the map. Although the nodes of the SOM
are distributed in a 2D grid, the boundaries of the grid are periodic, such
that the 2D projection is effectively an unravelled toroid. Wrapping the
boundaries ensures that trained nodes are not `pushed off' the boundaries of
the map. A plot of the 2D grid coloured by the value of the $i^{\rm th}$
weight of each node is called a component plane, and comparison of different
component planes can be used to study relationships between parameters in the
training set. 

Restricting the learning rate of the SOM as a function of time, and only
allowing it to change in ever finer regions, ensures that the introduction of
new training vectors refines the SOM, rather than obliterating the learning of
previous iterations. On this note, one requires the total learning time (i.e.
how many training vectors are used in the learning) to sufficiently
over-sample the input training set so that all training vectors are given a
chance to contribute to the learning at different stages of refinement. Note
that since the SOM is initialised randomly, and training vectors are selected
randomly, SOMs trained on the same input set will not `look' identical,
however the encoding of the map should be equivalent -- all that matters is
that similar nodes are close to each other (and distant from dissimilar nodes)
on the toroidal surface. The key characteristic of the self organisation is
that it retains the `topology' of the input training set, revealing
correlations between inputs that are not obvious. In fact, the SOM is often
described as a form of non-linear principle component analysis.

\subsection{A toy example}

Before we move to real world data, to demonstrate the concept of
self-organisation, we consider a simple toy example. In this example, we have
two `populations' which are simply represented as two gaussian distributions.
We will label these as `Red' and `Blue'. Red and Blue have means of $\mu_{\rm
Red}=1$, $\mu_{\rm Blue}=-1$ and both have scales $\sigma=0.5$. We now
randomly draw 10\,000 samples from Red and Blue and consider these as our
training set -- simply a list of 20\,000 numbers. Can we use self-organisation to separate these two
populations and predict whether a new test value belongs to the Red or Blue
population? Of course, this is a trivial example, because we could have
achieved the same result by simply plotting a histogram of the paramater
values, found the form of the toy distributions and therefore assign a
probability to any new value to determine the likelihood that it belongs to
Red or Blue. Still, this is a good demonstrative example.

We create a $20\times20$ node SOM, initialised with random weights selected
uniformly from the 20\,000 member training set. We set-up the initial SOM
parameters as described in \S2.1, and allow the total number of iterations to
be 200\,000, thus over-sampling the training set by a factor of ten. The
single component plane of the SOM seen at different stages of the learning
process is shown in Figure\,1. The component plane is coloured by
the `value' of the (in this case single element) weight of each node, each of
which has competed to represent the values of the training set. It is clear
that the map has arranged itself into two distinct regions -- this is because
the training sample is itself distributed around two distinct values (the
means of the distributions). Nodes that are close together (in terms of map
distance) have similar values, and there is a clear interface region on the
map where the distributions overlap. To reinforce this point, next to the
component plane we also show the so-called Unified Distance Matrix (UDM, or
U-Matrix), which visualises the average distance between the weights of
neighbouring nodes. The U-Matrix is a means of identifying boundaries of
`clusters' within the map. Small UDM values indicate that neighbouring nodes
have very similar weights and larger values indicate transition regions
between clusters.

The segregation of the different parts of the map allows us to label certain
nodes in the map as `Blue' and some as `Red' -- in other words, it will allow
us to classify new inputs (i.e.\ new samples that the SOM has never seen)
based on their BMU (it will either be in the Red or Blue class). To further
demonstrate this, in Figure\,1 we show two versions of the SOM but
this time scale the size of nodes based on the probability that their weight
value was drawn from the Red or Blue distributions. This clearly highlights
how the different parts of the map defined by the UDM correspond to the two
clusters in the input parameter space. In Figure\ 2 we show
the actual input distribution of Red and Blue objects, and the distribution of
the values of the weights of nodes in the trained map. Note that the SOM has identified several nodes which define an ambiguous classification where the two abundance of the two populations is equal at values near zero.

We have labelled 101/400 nodes as `Red' and 135/400 nodes as `Blue'
classifications based on the map division made apparent by the UDM. To test
the SOM, we use these nodes to classify 1000 {\it new} inputs from each of the
Red and Blue populations to find the identification rate, defined by the
fraction of new Blue or Red objects that correctly classified based on their
Best Matching Unit in the trained SOM. The results are shown in
Figure\,2. Not only does this simple classification procedure
successfully identify new test data, it correctly recovers the main properties
of the underlying distribution: the mean and standard deviations of the values
of objects classified as Red and Blue are $\mu_{\rm Red}=1.03$, $\mu_{\rm
Blue}=-1.01$, $\sigma_{\rm Red}=0.49$ and $\sigma_{\rm Blue}=0.51$, compared
to the input distribution of $\mu_{\rm Red}=1$ and $\mu_{\rm Blue}=-1$ and
$\sigma=0.5$. Exactly the same principle can be applied to astronomical data
sets, and so building on this trivial example we now demonstrate two real
world applications of a SOM trained on galaxy data from the COSMOS survey,
where we now include many more parameters in the training.

\begin{figure*} \vspace{-1.3cm}
\centerline{\includegraphics[scale=0.7,angle=-90]{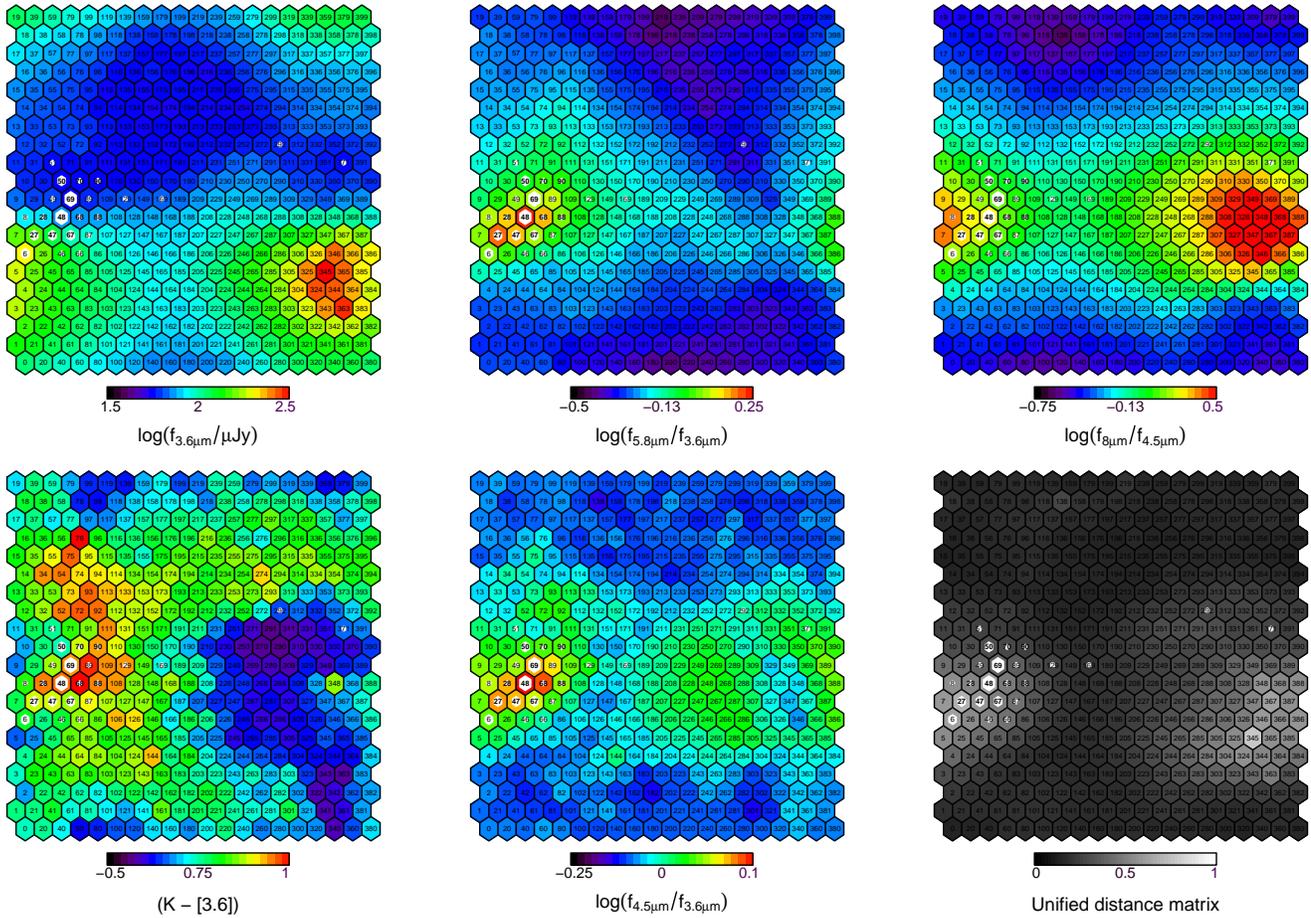}}
\caption{The component planes of a SOM trained on $\sim $$10^4$ IRAC-selected
galaxies in the COSMOS field (\S3.1). Note that the boundaries of each plane
are periodic. Nodes have been labelled with a numeric index to ease the
comparison between the component planes. The SOM was trained using the
5.8/3.6$\mu$m and 8/4.5$\mu$m colours and the 3.6$\mu$m flux density (top
panels), whereas the $K_s-{\rm [3.6]}$ and 4.5/3.6$\mu$m colours were phantom
vectors that did not contribute to the learning, but were allowed to follow
the SOM adaption. Several structures are visible that represent real
populations: for example the peak in the 3.6$\mu$m flux density map represents
stars, and there is clear segregation of red and blue IRAC sources that can be
matched to structures in the classic colour-colour plot. The red locus in the
5.8/3.6$\mu$m and 8/4.5$\mu$m colours can be identified, and we indicate the
positions of the BMUs of 83 spectroscopically identified BLAGN from {\it
z}COSMOS as white hexagons (scaled in size to represent the number of BLAGN
falling in each node). About half of the 83 BLAGN occupy just two nodes (48
and 69), and this helps us identify which nodes are best for our `AGN
selection'. The identification of clusters of nodes for object selection is
equivalent to colour-cuts in the traditional colour-colour plane, and we
illustrate this in Fig\,4.}\label{fig:som}\end{figure*}

\begin{figure*}
\centerline{\includegraphics[height=0.99\textwidth,angle=-90]{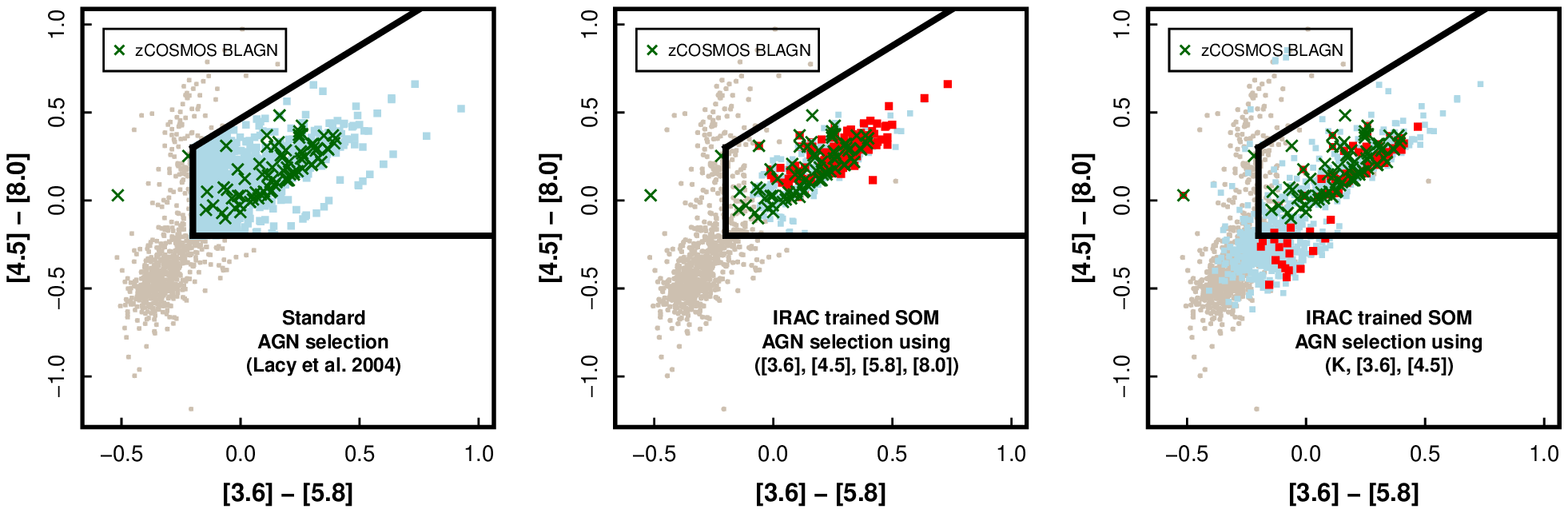}}
\caption{Traditional colour-colour plots for AGN selection. The first (left)
panel shows the standard wedge colour selection defined by Lacy et al.\ 2004
(see also Stern et al.\ 2005). We also show the position of spectroscopically
identified BLAGN from {\it z}COSMOS, highlighting the prominent colour-colour
locus. The central panel shows the galaxies selected as AGN using the nodes
identified in Fig\,3, where the BMU was calculated using both IRAC
colours. Red points represent our `robust' selection, utilising just two nodes
that best match $\sim$50\% of the 83 BLAGN (Fig\,3). Blue points
represent less secure classifications from a further 9 nodes. Together, these
identify 80\% of the known BLAGN. The final (right) panel shows the same SOM
selection, but this time the BMU was calculated using {\it only} the
$K_s$--3.6$\mu$m--4.5$\mu$m photometry. This demonstrates that the SOM can be
used as an effective selection tool, even when key information required for
the traditional selection is incomplete. In this case the 11 `AGN nodes'
correctly identified $\sim$40\% of the known BLAGN, but there is clearly some
scatter away from the locus. In both cases, the completeness could be improved
by including more nodes in the selection, but at the cost of
contamination.}\label{fig:agn} \end{figure*}

\section{Demonstrations using real data}

\subsection{Object classification and selection}

Colour-colour plots are a traditional method of isolating objects of interest,
since populations with similar spectral properties will have similar broadband
colours and therefore cluster together, or follow loci in appropriate
colour-magnitude or colour-colour planes. Perhaps the most successful example
in extragalactic studies is the selection of distant galaxies by virtue of the
Lyman break drop out, where UV--optical broadband filters that straddle the
redshifted Lyman break can efficiently sift $z\sim3$ galaxies from a field
(Steidel \& Hamilton\ 1993; Madau\ 1995; Steidel et al.\ 1996). There are many
similar examples of highly effective selection of objects using simple colour
criteria, and more recently this has been applied with great success for very
high-$z$ ($z\sim6$--$9$) galaxies that drop-out of optical bands altogether
(e.g.\ Bouwens et al.\ 2010; McLure et al.\ 2010). More complicated selections
can be constructed not only to pick-out galaxies at specific redshifts, but
also isolate those with certain properties (e.g.\ the star-forming / passive
$z>1.4$ galaxy selection of Daddi et al.\ 2004).

Here we demonstrate how the SOM can be used like an {\it N}-dimensional
colour-magnitude diagram, and when trained using a large catalogue, exploited
to identify those `clusters' of interesting objects. The trained SOM can then
be applied as a classification and filtering tool to extract objects of
interest from a new input catalogue.

\subsubsection{Selecting active galactic nuclei using {\it Spitzer} IRAC
colours}

{\it Spitzer} IRAC (3.6--8$\mu$m) colours have been shown to be very effective
at selecting AGN, including those whose optical emission is obscured by dust,
since these objects have a characteristic red power-law continuum in the
near/mid-infared that starts to dominate over the stellar emission at a
rest-frame wavelength of $>$2$\mu$m (Lacy et al.\ 2004; Stern et al.\ 2005).
This results in a red locus in IRAC colour space that stands out prominently
from the general galaxy population. Can we identify this population by
self-organising a catalogue of galaxies with IRAC photometry?

We take the {\it Spitzer} component of COSMOS (S-COSMOS, Sanders et al.\ 2007)
and set the training weights to be the 3.6$\mu$m flux and the
5.8$\mu$m/3.6$\mu$m and 8.0$\mu$m/4.5$\mu$m colours. A SOM is initialised with
$20\times20$ nodes. For the purposes of this demonstration we restrict the
catalogue to detections in all four bands and a 3.6$\mu$m flux limit of
50$\mu$Jy. The training catalogue has 10488 objects, and we allow the SOM to
iterate 104880 times in order to over-sample the catalogue by a factor of ten
during the training. The three component planes of the trained SOM are shown
in Figure\ 3, clearly showing the structures representing the
input catalogue (recall that the boundaries of the grid are periodic). To help
interpret the Figure, the nodes have been labelled with a numeric index, and
to understand the correlations between the component planes, one should
compare the value of common nodes in each plane. Correlations between the two
colours is apparent, and we can clearly identify the cluster of nodes that are
red in both sets of colours; these nodes represent the locus of galaxies with
power-law colours that would be apparent in the traditional colour-colour plot
(Fig.\,4). Therefore, rather than parameterising the AGN selection
as a series of cuts in colour space, we can simply use the map as a filter,
classifying any input galaxy as an AGN if its BMU is one of these nodes.

To illustrate the accuracy of the selection, and to verify which are the
correct nodes to use as robust `AGN selectors', we have taken the 83 galaxies
in the {\it z}COSMOS catalogue (Lilly et al.\ 2007) identified as Broad Line
AGN (BLAGN)\footnote{{\it z}COSMOS catalogue entries given the tag `13.x' or
`14.x'} and found their BMU in the trained SOM. Nearly 50\% of the BLAGN fall
in just two nodes, and 80\% are described by 11 nodes, most of which are
contiguous; we highlight these in Figure\,3. Although the SOM can
`discover' new classifications, the use of known objects (in this case
spectroscopically identified BLAGN) can be of great use when trying to label
nodes, and to assess the quality of subsequent selections. For example, the
two nodes that successfully describe 50\% of the BLAGN could be taken as
`high-confidence' AGN nodes, with the remaining nodes being lower confidence
selectors. Of course, as in normal techniques, there is a balance between
completeness and contamination. As a guide to the contamination rate, we
consider the two `high confidence' nodes and find out how many of the {\it
z}COSMOS galaxies that are not classified as BLAGN (and have very secure
redshifts, confidence class 3.5 or 4.5) fall into these nodes. Together, the
two nodes 48 and 69 pick out 114 galaxies that are not BLAGN, in addition to
the 39 (in the same redshift confidence class) that are. However, most of this
contamination is from just one node (69); if we restrict our AGN selection
node to 48 only (actually the reddest in IRAC colour, see Fig\,3),
of the 19 {\it z}COSMOS galaxies that match this node, {\it only one} is not
classified as a BLAGN. These contamination rates should only be taken as a
guide, given the likely incompletenesses in the spectroscopic selection and
classification of galaxies in the input {\it z}COSMOS catalogue. Nevertheless,
it is clear that the SOM could provide a very clean method for selecting
objects of interest.

In Figure\ 4 we plot the traditional IRAC colour-colour plane,
with the standard Lacy--Stern selection wedge indicated (although the Stern
selection is actually defined slightly differently, the broad selection is
effectively the same). As described above, we have chosen 11 nodes as our `AGN
classification', two of which we define as high-confidence. We then re-pass
the input catalogue through the SOM, this time noting which sources have BMUs
matching one of these classification nodes. The result is a clean selection of
galaxies along the expected AGN locus, and as expected we identify 80\% of the
BLAGN from the {\it z}COSMOS sample. It should be possible to refine the
efficacy of the selection by moving to a higher resolution SOM (more nodes),
which would improve the ability of the SOM to resolve finer details in the
topology of the data-set (in fact it is not clear what the optimum SOM
resolution is for a given training set, but ideally it should have many more
nodes than there are parameters, see \S3.3). Here we chose a fairly coarse
$20\times20$ SOM to better illustrate the component planes, and even at this
low resolution, the SOM is a remarkably powerful and clean selection tool.

To improve the selection of different types of AGN, more information could be
added to the training. For example, if one wanted to distinguish between
obscured and unobscured AGN, an optical band could be introduced: the
optical--NIR colours of obscured AGN are significantly redder than unobscured
AGN (e.g.\ Hickox et al.\ 2007). The extra information carried by, say, the
$(R-{\rm [4.5]})$ colour would allow the SOM to separate the two classes.

\begin{figure*}
\centerline{\includegraphics[height=0.99\textwidth,angle=-90]{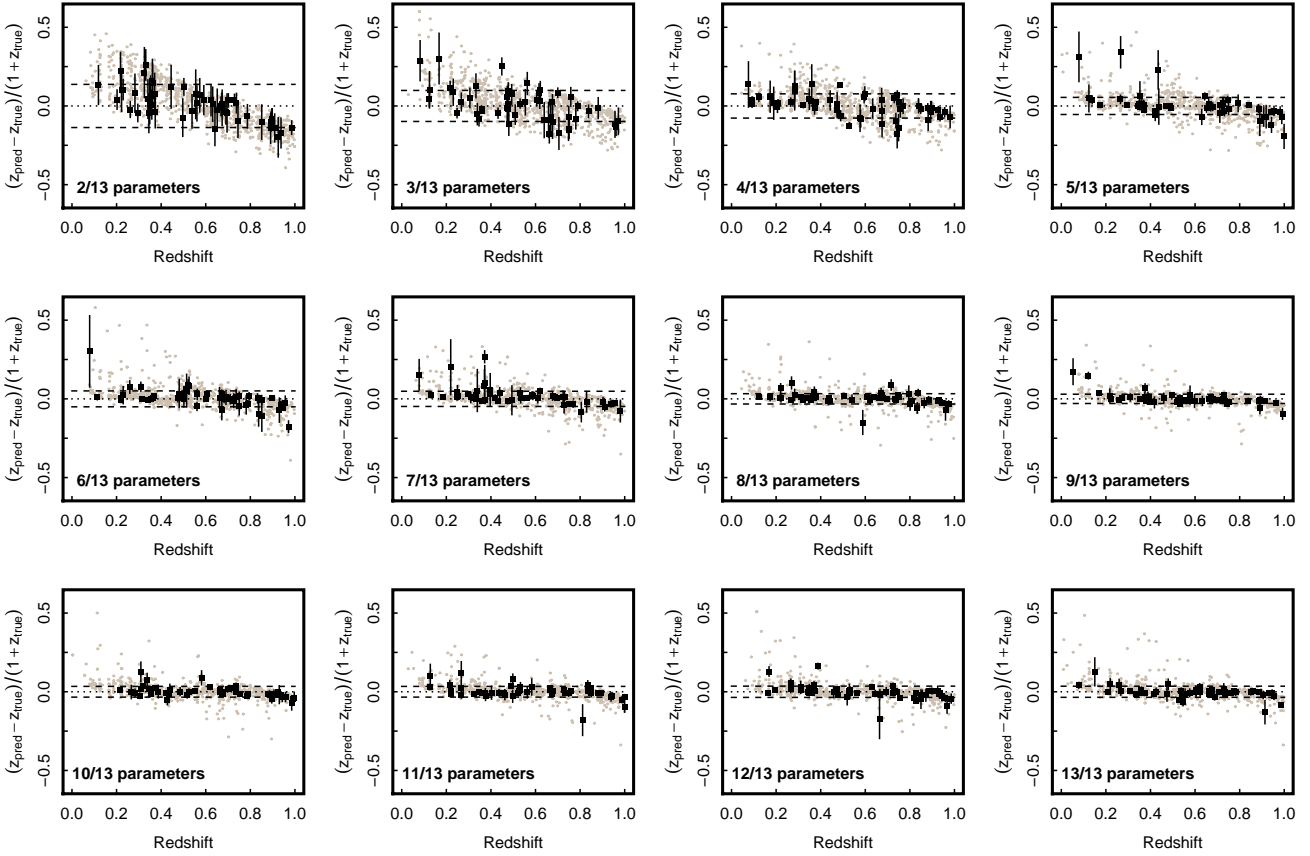}}
\caption{Photometric redshift accuracy recovered from a committee of 10 SOMs
trained on 3825 galaxies from the {\it z}COSMOS survey. Here we present the
predictions for another 3825 galaxies from the catalogue that {\it did not}
participate in the training (a randomly selected sub-set of 1000 points are
plotted for clarity). In each panel we highlight 50 randomly selected galaxies
and show their uncertainties, determined by the standard deviation of the
predictions made by the committee. The progression of panels row-wise from the
top left to bottom right shows the improvement of performance when
increasingly complete subsets of photometry for each galaxy. The first panel
just uses $u^*$-band and $r$-band monochromatic fluxes. The second panel
introduces the colour $(u^*-B)$, and subsequent panels include more colours up
to $({\rm [5.8]-[8.0]})$. The bias affecting predictions in the tails of the
redshift distribution can clearly be seen, and interestingly the accuracy
tends to asymptote after 9 parameters -- that is, all bands up to $K_s$ --
have been used. The projection of these plots as histograms is shown in Fig.\
6}\label{fig:zphot} \end{figure*}

In this special case we already knew that we were classifying AGN, and could
easily identify the part of the SOM that mapped this sub-population; a task
that was made easier with the sample of robustly identified BLAGN. However,
perhaps the most exciting possibility the SOM offers is the opportunity to
detect {\it new} classifications based on clustering in the parameter volume
that become apparent in component planes that would otherwise be undetected
using standard techniques. As described in \S2.3, one technique of identifying
significant clustering is to calculate the so-called Unified Distance Matrix
(UDM) or U-Matrix, which visualises the average `distance' to neighbouring
nodes. Clusters of nodes that are close to each other (i.e.\ similar UDM
values), bordered by regions where the UDM values are large could be
considered as clusters, and therefore potentially new classifications. We show
the UDM for the present example in Figure\,3, although this is not
always an appropriate method of identifying clusters. Upcoming large surveys
hold great promise for this type of data exploration; once new potential
classifications are identified with the SOM, it should be possible to isolate
those objects and properly assess their nature. The SOM provides a way of
finding those key, potentially rare objects from the overwhelmingly large
catalogues that are currently being produced.

\begin{figure*}
\centerline{\includegraphics[height=0.99\textwidth,angle=-90]{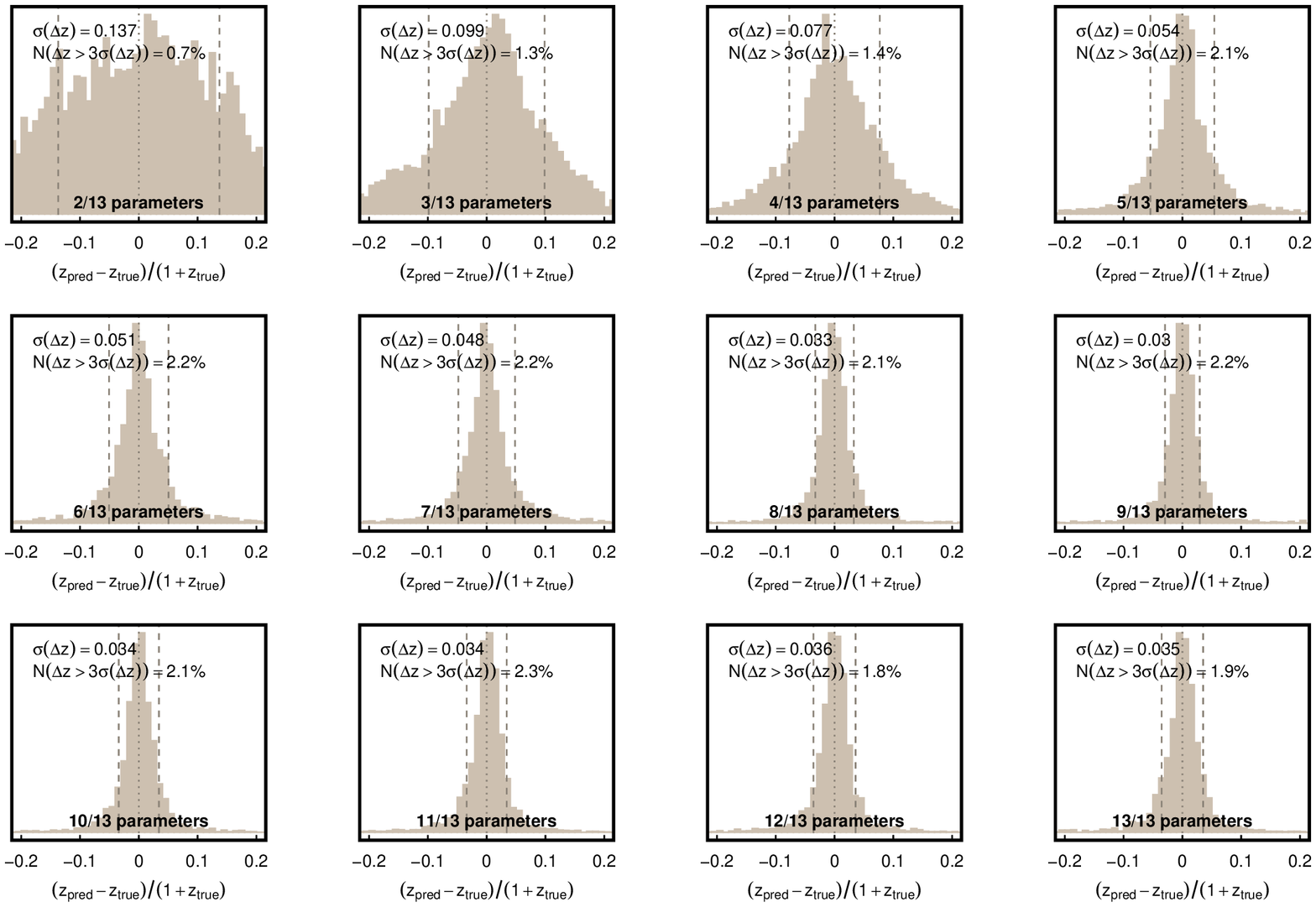}}
\caption{Distributions of redshift accuracy for the data shown in Fig\
5. The vertical lines indicate the r.m.s. value which we take to
be the figure of merit $\sigma(\Delta z)$. Surprising accuracy can be achieved
with a handful of parameters, and the improvement in accuracy reaches
$\sigma(\Delta z)\simeq0.03$ after 9 parameters are used, after rejecting a a
small outlier fraction of $\sim$2\%. This suggests that most of the
information driving the predictions, and encoded in the SOM are in the
optical--near-IR bands; inclusion of the IRAC bands for this sample, which is
limited to $z<1$, does not significantly improve the
accuracy.}\label{fig:zphot_hist} \end{figure*}

\subsubsection{Exploiting the map: the case of incomplete data}

What if we did not have the full slew of IRAC photometry for a new test galaxy
that we wish to classify, but instead have another photometric band? Here we
consider how cleanly the SOM trained above can select those obscured AGN using
just $K_s$, 3.6$\mu$m and 4.5$\mu$m photometry (this is a practical example,
now that {\it Spitzer} is operating in post-cryogenic `Warm mode' it has lost
the capability of its longer wavelength detectors, and so new fields will not
have 5.8 and 8$\mu$m photometry to perform the classic selection).

To approach this challenge, during the training of the SOM we allowed two
`phantom' components to be added to each training vector: $(K_s-{\rm [3.6]})$
and $([3.6]-[4.5])$. The $K_s$-band data (from the Infrared Side Port Imager
on the 4\,m Cerro Tololo Inter-American Observatory and FLAMINGOS on the 4\,m
Kitt Peak National Observatory) is taken from the COSMOS photometric catalogue
(2006 version, Capak et al.\ 2007). These additional components are not
allowed to take part in the learning (i.e.\ they are not considered in part 3
of the algorithm in \S2.2), but their values are still allowed to change, and
thus they get mapped into the SOM. In effect, this tells us what set of
$(K_s-{\rm [3.6]})$ and $([3.6]-[4.5])$ values correspond to the full-band
IRAC AGN selection described above. We can now introduce new test galaxies and
find out what their BMU is on the basis of {\it just} their $(K_s-{\rm
[3.6]})$ and $([3.6]-[4.5])$ colours. Again, if they match any of the nodes we
tagged as `AGN' above, these galaxies can be sifted out, but we expect the
selection to be less efficient, since it is now easier for galaxies to be
scattered away from the selector nodes. We plot the result of this alternative
selection in Figure\ 4. As expected, the efficacy of the selection
has been reduced, with only $\sim$40\% of the BLAGN identified, and more
scatter away from the standard locus (especially in the case of the
lower-confidence nodes). Nevertheless, the SOM is still effective at
picking-out AGN, even in this case where we are `missing' some of the
information used in the traditional selection.

As mentioned above, the performance could be improved by moving to a SOM with
a larger number of nodes, and thus allowing finer mapping resolution. In the
case of using $(K_s-{\rm [3.6]})$ and $([3.6]-[4.5])$ colours, contamination
could also be reduced by only using a sub-set of the nodes we have classified
as AGN on the basis of their 3.6--8$\mu$m colours -- i.e. just using the
`highest quality' nodes, as indicated in Fig\,4.

\subsection{Photometric redshifts}

\subsubsection{Setting up the problem}

When estimating a photometric redshift, we assume that there is a mapping
between a galaxy's true redshift $z$, and photometry vector $\mathbf{p}$ such
that $z = F({\mathbf p})$. If such a mapping exists, then the information to
find $F$ should be latent in a large galaxy catalogue where both photometry
and spectroscopic redshifts are available. Self-organisation of such a
training set will naturally encode $F$, and therefore a SOM can be used to
predict the redshifts (or indeed any other parameter that was involved in the
training) of new galaxies where, for example, only a subset of photometry is
known. This technique could be easily applied to a large imaging survey that
contains a smaller spectroscopic component in order to robustly estimate
redshifts for those galaxies lacking spectroscopic coverage. The advantage of
using a SOM for photometric redshift estimation is that it is completely
empirical, requires no assumptions about the spectral properties of the
galaxies and involves no user intervention to guide the learning (i.e.\ the
learning is unsupervised). However, there are two fundamental limitations to
the method:

\begin{enumerate}

\item[1)] The SOM cannot accurately extrapolate the properties of objects,
should they fall outside of the parameter volume of the original training set.
For example, if a catalogue limited to $z\leq 1$ is used to train the SOM, it
will catastrophically fail to predict the redshift of a $z=2$ galaxy, because
a galaxy of this type has not been `seen' by the SOM. Instances of such
failures could be flagged, because their `distance' from the BMU will be
large. It is therefore essential that the training set is a representative
sample, the larger the better, with a well known redshift distribution that
can aid in the interpretation of the reliability of predictions.

Actually, one {\it could} use information stored in the SOM to extrapolate
photometric redshifts beyond the range of the training set, in the sense that
the photometric weights of each node actually represent low-resolution
versions of the spectral shapes of galaxies (the broad-band photometry could
be simply interpolated to provide a continuous spectrum). Although more
time-consuming, if these low-resolution spectra were trusted to be a
representative sample of the full range of galaxy types, it should be possible
to use them in the usual $\chi^2$ template fitting procedures of other
photo-$z$ methods, allowing the spectra to redshift beyond the upper bound of
the training set and convolving with the relevant filter transmissions.

\item[2)] Related to (1), any biases in the training set will also bias the
prediction of unknown parameters in new test data. In this example that bias
might be the redshift distribution of the training set; the SOM will have seen
more examples of galaxies at the peak of the distribution compared to the
tails, potentially biasing the redshift estimates of galaxies in the tails
towards the centre of the distribution. Similarly, if the training set
contains exclusively red galaxies (classically selected Luminous Red Galaxies
for example), then the SOM will only be useful in predicting the properties of
input galaxies with similar characteristics. In summary, the predictive power
of the SOM is a strong function of the parameter distribution of the training
set, and so a proper understanding of the statistical nuances of the training
set is of critical importance when interpreting the SOM.

\end{enumerate}

In this demonstration, we again use the photometric data from COSMOS (Capak et
al.\ 2007) and S-COSMOS (Sanders et al.\ 2007), but merge it with 8910
spectroscopic redshifts from version 3.5 of the $z$COSMOS (bright) sample, the
$i<22.5$\,magnitude limited spectroscopic branch of the survey (Lilly et al.\
2007). The training sub-set is limited to galaxies with $z\leq1$ that are
detected in each of the {\it u$^*$BgVrizK$_s$} and [3.6], [4.5], [5.8], [8.0]
bands. In general, missing data (e.g.\ lack of coverage in a particular band
for a galaxy, perhaps due to masking or contamination) could be dealt with,
for example, by not allowing the missing weight to contribute to the learning
and/or handicapping the learning coefficient for that particular test vector.
Similarly, upper detection limits can be treated as equivalent to measurements
at the relevant significance, but for the purposes of clarity in this
demonstration, we require detection in all bands.

The number of objects in the catalogue after enforcing these constraints is
7651, with a median redshift of $z=0.55$. In order to test the predictive
power of the SOM, the sample is randomly split in two, such that one half of
the catalogue can be used for training, and the other for testing (where the
SOM has not had an opportunity to see those galaxies). The training set
therefore consists of 3825 unique inputs.

Multiple neural networks, or `committees', are often used to increase the
robustness of predictions (e.g.\ Collister \& Lahav\ 2004). Committees
introduce an extra level of stochasticity that provide a measure of
uncertainty through an examination of the fidelity of predictions made by
committee members. Here we initialise ten SOMs, each with $100$$\times$$100$
nodes, and set the total number of iterations per SOM to 382500, thus
over-sampling the input training set by a factor 10 for an individual map, and
a factor 100 over the committee. To introduce an extra level of randomness,
the initial choice of learning coefficient $L$ (equation\, 4) is selected from
a gaussian distribution centred at unity with a scale of 0.1; this allows each
SOM to learn at slightly different rates. The final predicted value is taken
to be the mean of the individual predictions from the committee members, and
the standard deviation of these predictions we take to be the uncertainty in the estimate. If
we had used many more SOMs in the committee, it should be possible to collect
the results together to form a probability density function for the parameter
prediction, which might provide a better representation of the uncertainty
(note that SOMs can be trained in parallel for this purpose).

In our example, for each training vector we have a set of broad band
photometry and a spectroscopic redshift. We set these as the weights of each
training vector. To reduce the parameter space, we assign the photometry as a
set of colours in consecutive bands, $(u^*-B)$, $(B-g)$, $(g-V)$, and so-on up
to ${\rm ([5.8]-[8.0])}$. We also include the single $u^*$ and $r$ magnitudes
as monochromatic flux measurements, and finally the spectroscopic redshift
from $z$COSMOS. In total, each training vector contains 14 elements. After
training all SOMs in the committee, we test the predictive power of the SOM
ensemble using the half of the catalogue that did not participate in the
training, calculating the BMU for each object using sub-sets of the photometry
(e.g.\ just $(u^*-B)$, then adding $(B-g)$, $(g-V)$, and so-on until we
include all photometry weights up to the IRAC bands). In this case, the
spectroscopic redshift component of the weight is not considered when
calculating the BMU. In each trial, the redshift weight tagged to the BMUs
provides the `photometric' redshift, and these are averaged over the committee
to give the final prediction. As we know what the true redshift of each test
galaxy is, we can assess the accuracy of the method.

\subsubsection{Photometric redshift accuracy}

We define the figure of merit for the photometric redshift accuracy in the
usual way as the root mean square of the difference between the true and
estimated redshift $\sigma(\Delta z) = \surd\left< \Delta z^2\right>$, where
$\Delta z = (z_{\rm spec} - z_{\rm phot}) / (1+z_{\rm spec})$. Figures\
5 and 6 shows the results, where we have
interrogated the committee of ten SOMs for the photometric redshift of a test
galaxy with increasingly complete sub-sets the full range of photometry. There
is a clear decline in $\sigma(\Delta z)$ as more photometric information
information is added, asymptoting at $\sigma(\Delta z)\sim0.03$. Surprising
accuracy can be achieved with a rather sparsely sampled input vector, however
in these cases one can clearly see the bias described above that results in
the overestimation of redshifts for galaxies at $z<\left< z\right>$ and vice
versa. Where shown, the error bars are the standard deviation of redshifts
recovered from the ten SOMs. This is certainly an underestimation of the true
error; one could also incorporate the formal photometric uncertainties by
running the SOM interrogation several times and allowing each photometry value
to randomly scatter about its mean according to its 1$\sigma$ measured
uncertainty. In this example, large error bars simply reflect cases where
galaxies with similar characteristics were poorly represented in the training
set, and thus are scattered between dissimilar BMUs in each committee member.

Using the $u^*$-band to $K_s$-band photometry, we can achieve $\sigma(\Delta
z)=0.03$ after rejecting $\sim$2\% $>$$3\sigma$ outliers. Including the IRAC
bands does not significantly improve the accuracy, despite the fact they were
included in the training: $\sigma(\Delta z)$ no longer improves after the 9th
parameter ($z-K_s$) is added. This reflects the fact that for $z<1$ it is the
$\lambda<2$$\mu$m photometry that `carries' most of the information required
for the photometric redshift (as expected; the 4000\AA\ and Balmer breaks are
still blueward of the $J$-band at $z<1$, and the 1.6$\mu$m stellar bump,
another good redshift discriminant is just redward of $K_s$).

This accuracy is comparable to, or rivals, that which can be achieved with
traditional spectral template fitting techniques. Pertinent to this data-set,
Mobasher et al.\ (2007) achieved $\sigma(\Delta z)=0.031$ with a template
fitting technique to 16 photometric bands in the COSMOS field. This was found
to be in good agreement with photo-{\it z}s derived from the independent codes
Le Phare (Arnouts et al.\ 1999), BPZ (Benitez\ 2000) and ZEBRA (Feldmann et
al.\ 2006). Several of these methods use Bayesian inference to derive
photometric redshifts. It should be noted that more recently Ilbert et al.\
(2009) achieved much higher photometric redshift accuracies $(\sigma(\Delta
z)<0.01)$ in the COSMOS field using the Le Phare code (S. Arnouts \& O.
Ilbert) with 30 broad-, medium- and narrow-bands for template fitting. Given
the improvement seen in the SOM photo-{\it z} technique when more photometric
bands are introduced, we would anticipate an improvement in our reported
accuracy if we re-trained the SOM with a similar large number of bands.

One of the main benefits of the SOM technique, aside from the non-reliance on
assumptions of spectral properties, is the speed at which photometric
redshifts can be calculated once training has completed. The time to calculate
the photometric redshift and error is simply the computational time to query
each SOM to find the BMU -- less than a few hundredths of a second per galaxy
on a typical modern desktop machine\footnote{The computations presented in
this paper were performed on a 3.2\,GHz Intel Core i3 iMac with 16\,GB RAM,
and the coding was certainly not optimal.}.

\subsubsection{Additional photometric redshift tests and comparisons: PHAT}

Hildebrandt et al.\ (2010) present a system for the consistent testing of
different photo-{\it z} codes: `PHAT: PHoto-{\it z} Accuracy
Testing'\footnote{http://www.astro.caltech.edu/twiki\_phat/bin/view/Main/WebHome}.
PHAT provides a standard mock catalogue containing galaxies represented by the
empirical spectral energy distribution templates of Coleman,\ Wu \& Weedman\
(1980) and Kinney et al.\ (1996), together covering the full range of galaxy
spectral types from passive ellipticals to starburst systems. Synthetic colour
information for each galaxy is calculated for each template for photometric
bands spanning the ultraviolet to mid-infrared, specifically: the
Canada-France-Hawaii Telescope MEGACAM {\it ugriz}-bands, the United Kingdom
Infrared Telescope {\it YJHK}-bands and the 3.6$\mu$m and 4.5$\mu$m {\it
Spitzer} IRAC bands.

As ours is an empirical method and requires a training set where the redshift
is known, we use the `large' PHAT catalogue of 170\,000 objects with noise
included (where a parametric model for the signal-to-noise ratio as a function
of source flux is used, and photometry perturbed accordingly according to a
gaussian distribution). We create a training sub-set by randomly sampling 10\%
of the full catalogue. In this example, we initialise a $200\times200$ SOM and
set the number of iterations to oversample the training set by a factor of 5.

Hildebrandt et al.\ (2010) define the photo-{\it z} accuracy figure of merit
as the mean and scatter (rms) in $\Delta z = z_{\rm model} - z_{\rm phot}$,
and the outlier rate as the fraction of objects with $|\Delta z|>0.1$. For
comparison with the results presented in Hildebrandt et al.\ (2010) for
PHAT-testing of 16 recent photo-{\it z} codes (several of which are in
widespread use), we calculate the same statistics on the predicted redshifts
retrieved for the galaxies that did not participate in the training of our
SOM. The best codes tested by Hildebrandt et al.\ (2010) typically have
$\left<|\Delta z|\right >\leq 0.005$, scatters of $\sigma(\Delta z)
\sim$0.01--0.02 and small outlier rates of $<0.1$\%. Testing the trained SOM
on a sub-sample of 100\,000 galaxies from the large catalogue that did not
participate in training we find an average $\left<\Delta z\right >
=-7\times10^{-4}$, $\sigma(\Delta z) = 0.016$ and outlier rate of 0.13\%. The
relatively large outlier rate (compared to some of the codes tested in
Hildebrandt et al.\ 2010) is driven by the poorer accuracy at the tails of the
redshift distribution, which is a natural bias in this method. When
considering only galaxies in the range $0.1<z<0.5$, although the rms accuracy
is the same, the outlier rate drops to 0.06\%. Thus, the empirical SOM method
for photo-{\it z} prediction is competitive with established photo-{\it z}
codes. It is likely that the accuracy could be improved further by using an
even larger training sample with a longer learning period, at the expense of computational time.

\subsection{Limiting factors}

Aside from the limitations discussed above regarding the choice training set,
and the natural biases that are encoded into the SOM, there are several other
important issues to consider, and we briefly review these here.

The rate of learning, or how quickly the SOM adapts during training, is set by
(a) two learning coefficients (equation 1 and 4) which vary as a function of
node distance and learning time; (b) the rate of decay of these coefficients;
(c) the shape of and rate of decay of the region of influence around the BMU
where neighbouring nodes are allowed to change; (d) the size, or resolution of
the SOM, and (e) the total learning time. It is not clear what the optimum
combination of these factors is that would produce the best mapping is, and it
would take a long time to do so. So, the exact choice of training parameters
might be the main limiting factor in the SOM technique. However, during the
course of this investigation, we have found some simple configurations that
appear to produce robust results.

First, the number of nodes in the SOM should be initialized such that the
total number scales roughly with the number of training parameters,
$\sim$$2^N$, and a good minimum is $\sim$400 nodes arranged in a $20\times20$
grid\footnote{Note that there is nothing to preclude arranging the nodes in a
3 dimensional (or higher) grid, but this would defeat the purpose of
visualisation.}. This is to allow the mapping to `resolve' possible
correlations and clustering between several parameters. Clearly, when making
predictions for new test data, the size of the SOM sets a fundamental limit on
the accuracy, as the total input parameter space is discretised into a finite
number of bins. In the case of this photometric redshift example, we set the
total number of nodes to be $10^4$, and this seems adequate to make accurate
predictions whilst keeping down training time. In the example of object
selection however, we were more interested in training the SOM to make
selections of objects in rather broad swathes of the parameter space, and so
in this case a SOM with fewer nodes was successful (and is also beneficial for
visualisation purposes). Note that there are variant SOM algorithms that allow
the number of nodes in the map to be dynamic, growing according to the need of
the training sample (Alahakoon \& Halgamuge\ 1998)

The total number of iterations was set to ten times the number of elements in
the training set. This was to allow the SOM to see each training vector about
ten times, and participate in the refining of the self-organisation at
different stages in the learning process. We could envision even better
results if we allowed more over-sampling of the training set, but this comes
at the cost of longer training times.

Finally, we initialised the learning co-efficients to unity, and set the size
of the neighbourhood function to be approximately half the linear size of the
SOM. This initial size allows test vectors selected at the start of the
learning to influence large, unrefined sectors of the map. As iterations
cumulate, new training vectors simply refine the map, contributing less
drastic changes due to the decreasing size of the neighbourhood function and
declining learning coefficients. We found that the shape of the neighbourhood
function (a gaussian), the rate of its decay (the size decreases linearly with
time), and the decay of the learning coefficients produced excellent results
in multiple SOM realisations involving different types of data. Again, a
future study could investigate what the optimal learning parameters are, with
the best results perhaps coming from a more extended committee of hundreds or
thousands of SOMs (that could be trained in parallel), each with different
self-organisation styles and learning capabilities.

\section{Summary}

Self-organised maps (SOMs) are a class of neural network that employ
unsupervised learning to map the topology of a multidimensional data set. This
is a powerful method for exploring large catalogues of astronomical data; the
method can discover correlations between parameters, detect clustering within
the parameter volume, and can be exploited to predict the parameters of new
test data in a completely empirical way. Here we have presented the SOM as a
potential tool for current and future large astronomical surveys, highlighting
two practical examples:

\begin{enumerate}

\item Selection of galaxies with active nuclei trained on {\it Spitzer} IRAC
colours (3.6--8$\mu$m). The SOM trained on IRAC photometry naturally `finds'
the characteristic colours of obscured AGN, and the corresponding nodes can be
used as a filter to select similar objects from a new data set. This filter
can even be used where the information used in the training is incomplete, or
unavailable. We demonstrate that the same SOM can be used to select known AGN
using just $K_s$, [3.6] and [4.5] photometry. While we chose AGN as a
demonstrative example, SOMs could be used to select a wide range of
astronomical objects, with the exciting possibility that self-organisation
could discover `new' classifications in upcoming large data surveys.

\item Estimation of redshifts from broad-band photometry, trained using a deep
spectroscopic survey: $z$COSMOS. The accuracy of the redshift estimation
defined by the r.m.s. in $\Delta z = (z_{\rm pred}-z_{\rm true})/(1+z_{\rm
true})$ is $\sigma(\Delta z)=0.03$, with a small outlier rate of $\sim$2\%,
competitive with other established photo-$z$ codes using alternative
techniques for deriving the redshift from photometry. We also test the SOM as
a photo-{\it z} tool using the PHoto-{\it z} Accuracy Testing catalogue
(Hildebrandt et al.\ 2010), which provides a much larger training set with
model galaxies covering a range of spectral types, and 10 bands of broadband
photometry. We find that the photo-{\it z} accuracy of the SOM is competitive
with many established photo-{\it z} codes, delivering an rms in $(z_{\rm
true}-z_{\rm pred})=0.016$ with a small outlier rate of 0.13\%.

Accuracies could be significantly improved by training on a larger training
sample, but other factors also affect performance, including the `resolution'
of the SOM, the choice of learning coefficients, and so-on. Although not
without its limitations, which are discussed, the advantages of using a SOM
for predicting photometric redshifts (or any other parameter) are (a) it is a
completely empirical method; (b) once training has completed, predictions can
be achieved very quickly, since the only cost function that has to be
evaluated is the location of the best matching node (BMU) for a new test
galaxy. \end{enumerate}

We have demonstrated two simple examples here, using one of the most basic SOM
algorithms, but there are many practical applications beyond what has been
presented. One could envision more extravagant training scenarios,
applications for and adaptations to the algorithm that might prove fruitful.
In conclusion however, we suggest that SOMs are versatile tools that could be
used in data mining and visualisation applications for existing and up-coming
large surveys, where efficient techniques will be required to fully harness
the power of the exceptionally large and intertwined databases set to flood
the community. 

\section*{Acknowledgements}

J.E.G.\ is supported by a Banting Postdoctoral Fellowship, administered by the
Natural Science and Engineering Research Council of Canada. We thank the
anonymous referee for comments and suggestions that improved the paper, and
Kristen Coppin, Duncan Farrah, Ryan Hickox and Phil Marshall for helpful
comments and discussions.

This research has made use of the NASA/IPAC Infrared Science Archive, which is
operated by the Jet Propulsion Laboratory, California Institute of Technology,
under contract with the National Aeronautics and Space Administration

\section*{Nomenclature and notation}

\begin{tabular}{lp{0.275\textwidth}}
	
	SOM & Self-organised/organising map\cr
	\cr
	Node & Single `neuron' in SOM; nodes are arranged on the surface of a 3D toroid, but visualised unravelled, in 2D \cr
	\cr
	Training vector, $\bf t$ & Example of single data element from training set (e.g.\ galaxy photometry)\cr
	\cr
	Weight vector, $\bf w$ & Vector of identical size to ${\bf t}$ attached to each node that competes to become more like the training vector\cr
	\cr
	BMU & Best Matching Unit, is the `winning node' that is most like a randomly sampled training vector\cr
	\cr
	U-Matrix & Unified Distance Matrix (UDM: a method of visualising and detecting clustering in the map using the average distance to neighbouring nodes \cr
	\cr
	$r_{\rm BMU}$ & Radius of learning influence of BMU\cr
	\cr
	Neighbourhood function & Form of spatial learning function within $r_{\rm BMU}$\cr
	\cr
	Component plane & 2D representation of the values of the $i{^{\rm th}}$ element of the weight vector of nodes in the map\cr
	\cr
	Learning rates & Co-efficients determining the amount that weights can adapt to become more like training vectors; these vary spatially (relative to the BMU) and temporally, tending toward zero over the duration of training\cr
	\cr
	Over-sampling & Number of times a given training vectors are `seen' by the SOM during learning\cr 
	\cr
	Committee & Several SOMs trained independently on the same training set\cr
	
\end{tabular}


\label{lastpage}

\end{document}